\documentclass[aps,prl,superscriptaddress,twocolumn,nobibnotes]{revtex4-1}
\pdfoutput=1
\usepackage{amsmath}
\usepackage{ amssymb }
\usepackage{graphicx}
\usepackage{color}
\usepackage{bibunits}

\begin{document}
\begin{bibunit}[apsrev4-1]

\newcommand{\beq}{\begin{equation}}
\newcommand{\eeq}{\end{equation}}
\newcommand{\beqa}{\begin{eqnarray}}
\newcommand{\eeqa}{\end{eqnarray}}
\newcommand{\ben}{\begin{enumerate}}
\newcommand{\een}{\end{enumerate}}
\newcommand{\hs}{\hspace{0.5cm}}
\newcommand{\vs}{\vspace{0.5cm}}
\newcommand{\note}[1]{{\color{red} #1}}
\newcommand{\tim}{$\times$}
\newcommand{\bigo}{\mathcal{O}}
\newcommand{\bra}[1]{\ensuremath{\langle#1|}}
\newcommand{\ket}[1]{\ensuremath{|#1\rangle}}
\newcommand{\bracket}[2]{\ensuremath{\langle#1|#2\rangle}}

\title{Detecting Goldstone Modes with Entanglement Entropy}

\author{Bohdan Kulchytskyy}
\email{bkulchyt@uwaterloo.ca}
\affiliation{Department of Physics and Astronomy, University of Waterloo, Ontario, N2L 3G1, Canada}

\author{C. M. Herdman}
\affiliation{Department of Physics and Astronomy, University of Waterloo, Ontario, N2L 3G1, Canada}
\affiliation{Department of Chemistry, University of Waterloo, Waterloo,  Ontario, N2L 3G1, Canada}
\affiliation{Institute for Quantum Computing, Waterloo, Ontario, N2L 2Y5, Canada}

\author{Stephen Inglis}
\affiliation{ Department of Physics, Arnold Sommerfeld Center for Theoretical Physics and Center for NanoScience, 
University of Munich, Theresienstrasse 37, 80333 Munich, Germany}

\author{Roger G. Melko}
\affiliation{Department of Physics and Astronomy, University of Waterloo, Ontario, N2L 3G1, Canada}
\affiliation{Perimeter Institute for Theoretical Physics, Waterloo, Ontario, N2L 2Y5, Canada}

\date{\today}

\begin{abstract}
In the face of mounting numerical evidence, Metlitski and Grover [arXiv:1112.5166] have given compelling 
analytical arguments that systems with spontaneous broken continuous symmetry contain a sub-leading contribution
to the entanglement entropy that diverges logarithmically with system size.  They predict that the coefficient of this log
is a universal quantity that depends on the number of Goldstone modes.  In this paper, we
confirm the presence of this log term through quantum Monte Carlo calculations of the second R\'enyi entropy on the spin 1/2 XY model.  
Devising an algorithm to facilitate convergence of entropy data at extremely low
temperatures, we demonstrate that  
the single Goldstone mode in the ground state can be identified through the coefficient of the log term.  Furthermore,
our simulation accuracy allows us to obtain an additional {\it geometric} constant additive to 
the R\'enyi entropy, that matches a predicted fully-universal form obtained from a free bosonic field 
theory with no adjustable parameters.

\end{abstract}

\maketitle

{\em Introduction -- }
In condensed matter, the entanglement entropy of a bipartition contains an incredible amount of information
about the correlations in a system.  In spatial dimensions $d \ge 2$, quantum spins or bosons display an entanglement entropy
that, to leading order, scales as the boundary of the bipartition \cite{arealaw1,arealaw2,Eisert:2010}.  Subleading to this ``area-law'' are various
constants and -- particularly in gapless phases -- functions that depend non-trivially on length and energy scales.
Some of these subleading terms are known to act as informatic ``order parameters'' which can detect non-trivial correlations, such
as the topological entanglement entropy in a gapped spin liquid phase \cite{Alioscia1, Alioscia2, Kitaev:2006,Levin:2006}.  At a quantum critical point, subleading
terms contain novel quantities that identify the universality class, and potentially can provide constraints
on renormalization group flows to other nearby fixed points \cite{Fradkin:2006,Casini:2007,Myers:2012,Singh:2012t,Inglis_2013,Kallin:2013,Grover:2014}.

In systems with a continuous broken symmetry, evidence is mounting that the entanglement entropy between two
subsystems with a smooth spatial bipartition contains a term, subleading to the area law, that diverges logarithmically
with the subsystem size.  First observed in spin wave \cite{Song:2011} and finite-size lattice numerics \cite{Kallin:2011}, the apparently anomalous logarithm
had no rigorous explanation until 2011, when Metlitski and Grover developed a comprehensive theory \cite{Max_Tarun}.  They argued that, for a 
finite-size subsystem with length scale $L$, the term is a manifestation of the two long-wavelength energy scales 
corresponding to the spin wave gap, and the ``tower of states'' arising from the restoration of symmetry in a finite volume
\cite{Anderson:1952, Bernu:1992, Lhuillier:2005, Misguich:2007}.
Remarkably, their theory not only explains the subleading logarithm, but predicts that the coefficient is directly proportional to 
the number of Goldstone modes in the groundstate.
Furthermore, describing a Goldstone mode with a free scalar field theory allows them to predict the 
value for an additional additive {\it geometric} constant, which is fully universal and should therefore be the same across a wide range of continuum 
theories and lattice models.

\begin{figure}[t]
\begin{center}
\scalebox{1}{\includegraphics[width=0.8\columnwidth]{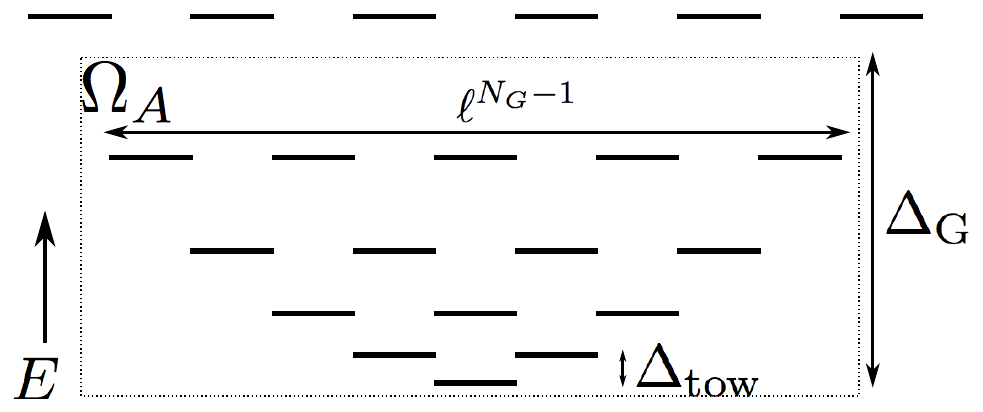}}
\end{center}
\caption{Schematic energy level structure of the low energy tower of states for finite-size systems with spontaneous breaking of a continuous symmetry. The  correction to the entanglement entropy may be approximated by the log of the number of quantum rotor states below the Goldstone gap, $\Omega_A$, which is represented by the states within the dotted box.}
\label{fig:tower}
 \end{figure}

In this paper, we confirm these predictions in a striking way, through large-scale quantum Monte Carlo (QMC) simulations
on a spin-1/2 XY model on the square lattice.  By employing an extended-ensemble generalization of a {\it ratio method} \cite{Gelman:1998, Humeniuk:2012, Luitz:2014}, we are able to 
carefully converge the mutual information of the second R\'enyi entropy to its low-temperature value.  
There, finite-size scaling reveals the coefficient of the subleading logarithmic term to precisely match the 
prediction of Metlitski and Grover, identifying the lone Goldstone mode in the theory.  Remarkably, our simulations are
accurate enough to also measure the universal additive geometric constant, which is consistent with the predicted
relationship to the constant which appears in a lattice-regularized free scalar field theory \cite{Max_Tarun}.

{\em Entanglement entropy in the tower of states -- }
To obtain a qualitative understanding of the origin of the logarithmic correction in Metlitski and Grover's theory, 
it is simplest to first envision decoupling the two spatial subsystems, $A$ and $B$ , that define the entangled bipartition. 
The low-energy degrees of freedom in each subsystem can be described by an $O(N)$ rotor ($N=2$ for the XY model), 
representing the direction of the order parameter.
Here we are only allowing global fluctuations of the order parameter within each subsystem, such that we may approximate the state of $A$ and $B$ each as a single independent quantum rotor.
The effective Hamiltonian
of each subsystem is  $H =  {\bf L}^2 /2 I$, where ${\bf L}^2$ is the total angular momentum operator with eigenvalues $\ell(\ell+1)$ and $I$ is the effective moment of inertia which is extensive, proportional to the magnetic susceptibility $\chi$: $I \sim \chi L^d$ in $d$ spatial dimensions \cite{Sandvik:2010}. Thus the energy scale of the tower $\Delta_{\rm{tow}} = 1/\chi L^d$ vanishes with the system volume, faster than any other energy scale.
The eigenstates of ${\bf L}^2$ result in the famous ``tower of states'' observed routinely in computational studies of systems with continuous
symmetry breaking in a finite volume~\cite{Bernu:1992, Lhuillier:2005, Misguich:2007}. 

The interaction between $A$ and $B$ which aligns the subsystem order parameters may be introduced via a Goldstone mode Hamiltonian $H_G$ which couples the two rotors. The energy scale of the $H_G$ is the Goldstone mode gap $\Delta_G$ which is the 
scale of the lowest energy spin waves. Since $\Delta_G \sim c/L$ where $c$ is the spin-wave velocity, $\Delta_G \gg \Delta_{\rm{tow}}$ in the thermodynamic limit for $d>1$. In the limit $\Delta_G\rightarrow \infty$, there are no \emph{relative} fluctuations in the order parameter between subsystems, and $A$ and $B$ act as a single rigid rotor. For finite $\Delta_G$, there will be relative fluctuations between the subsystems order parameters due to the zero point fluctuations of $H_G$.

To estimate the entanglement entropy contribution from the tower of states, we can count the number of ``accessible" states of subsystem $A$, $\Omega_A$, when the total system is in the ground state, and use $S_{\rm{tow}} \sim \log \Omega_A$. 
In the limit $\Delta_{G}\rightarrow \infty$ and the rotors are rigidly coupled, the ground state is the ground state of the total system tower of states with zero total angular momentum: $\ell_{\rm{AB}}=0$. In this case all states in the $A$ subsystem tower 
are accessible to the ground state, as each state in $A$ can be paired with an appropriate state in $B$ to form a state with nonzero overlap with
the $\ell_{\rm{AB}}=0$ state. However, as discussed above, by including a \emph{finite} $\Delta_{G}$ and thus allowing \emph{relative} fluctuations 
of the subsystem order parameter between $A$ and $B$, the fluctuations in the subsystem angular momentum are finite and determined by the ratio
 of the energy scales: $\langle {\bf L}_A^2 \rangle \sim \Delta_G/\Delta_{\rm{tow}}$ \cite{Max_Tarun}. 
In fact, the reduced matrix of the subsystem takes the form of a thermal density matrix with an effective ``entanglement Hamiltonian'' given by $H_{\rm{tow}}$ and the ``entanglement temperature'' given by $\Delta_G$~\cite{Max_Tarun}; the resulting tower of states structure in the entanglement spectrum has been seen in numerics~\cite{Kolley:2013,Alba:2013,Lauchli:2013,Poilblanc:2014}. 
Thus the inclusion of Goldstone modes cuts off the accessible states of the subsystem to those with an energy below the spin wave gap, as illustrated in Fig. \ref{fig:tower}.
 
As an example of this mechanism, consider the case of $N=2$ (valid for our XY model simulations below). Here the rotors have a single component of angular momentum $\ell^z$ and the orientation of the rotors is described by a single angle $\theta$. For $\Delta_G \rightarrow \infty$ the ground state has $\ell^z_{\rm{AB}} = 0$, which has nonzero overlap with states of equal and opposite $\ell^z$ in each subsystem: $\ket{\ell^z_A = \ell,\ell^z_B = -\ell}$; consequently all $\ket{\ell^z_A}$ state are accessible in this limit. We may include the effect of the lowest Goldstone mode by treating the dynamics of the relative angle between subsystems $\theta_\delta$ as a single harmonic oscillator with frequency $\Delta_G$ and moment inertia $I_{\delta} \sim \Delta_{\rm{tow}}^{-1}$, with an effective Hamiltonian
\begin{equation}
H_G = \frac{1}{2I_\delta} L^2_\delta + \frac{1}{2} I_\delta \Delta_G^2 \theta_\delta^2 .
\end{equation}
Here, the fluctuations in the relative angular momentum $L_\delta$ are given by the ground state fluctuations of a harmonic oscillator: $\langle L_\delta^2 \rangle \sim I_\delta \Delta_{\rm{G}}/2 \sim \Delta_{\rm{G}}/\Delta_{\rm{tow}}$. The key point here is that because the order parameter is canonically conjugate to the rotor angular momentum, \emph{increasing} the relative fluctuations in the order parameter \emph{reduces} the fluctuations in $L^2$. Thus, allowing relative fluctuations of the order parameter between subsystems effectively cuts off subsystem rotor states that are accessed in the ground state at order $\ell \sim (\Delta_G/\Delta_{\rm{tow}})^{1/2}$
 -- a relationship that holds for all $N$ \cite{Max_Tarun}. 
   
We may therefore estimate $\Omega_A$ by counting the number of states (in $A$'s tower of states) that lie below $\Delta_G$. For systems with $O(N)$ symmetry,
 the tower of states is described by a rotor living on an $N_{G}=N-1$ dimensional sphere, where $N_G$ is the number of Goldstone modes. The dengeneracy
 of each energy level is of order $\ell^{N_G-1}$. We then may estimate the total number of states below $\Delta_G$ by integrating the degeneracy up to the 
 cutoff $\ell_{\rm{co}} = (\Delta_G/\Delta_{\rm{tow}})^{1/2}$:
\begin{equation}
\Omega_A \sim \int_0^{\ell_{\rm{co}}} d \ell~ \ell^{N_G-1} \sim \left( \frac{\Delta_G}{\Delta_{\rm{tow}}} \right)^{N_G/2}.
\end{equation}
Using the relation $\chi = \rho_s/c^2 $ from hydrodynamic spin-wave theory where $\rho_s$ the stiffness~\cite{HALPERIN1969}, the entanglement entropy correction due to the tower of states becomes
\begin{equation}
S_{\rm{tow}} \sim \frac{N_G}{2} \log \left( \frac{\rho_s}{c} L^{d-1} \right).
\end{equation}
We see that the logarithmic correction to the area law arises due to the quasi-degeneracy of accessible bulk subsystem states, that scales as a 
power law in $L$ for systems with spontaneously broken continuous symmetries.  This contrasts with the leading-order area law, arising from the exponential
scaling of the number of local boundary states with the boundary area.  Clearly, the prefactor of the logarithmic correction is a universal number that simply counts the number of Goldstone modes.

{\em Quantum Monte Carlo Procedure -- }
In order to examine the effects of broken continuous symmetry on entanglement, we implement a highly-efficient Stochastic 
Series Expansion (SSE) QMC 
algorithm \cite{SSE1,SSE2,Sandvik:1999} for the 2D spin-1/2 XY model, $H = J \sum_{\langle i j \rangle} (S^x_i S^x_i + S^y_i S^y_i )$.  
This model is known to realize a ground state where the $U(1)$ symmetry is spontaneously broken, resulting
in one Goldstone mode. The finite-temperature SSE algorithm uses
a version of the directed-loop updates specialized to the XY model \cite{Dir_loop}.  To measure the entanglement entropy, we employ a
replicated simulation cell $Z[A,2,T]$, which gives access to second R\'enyi entropy 
$S_n = \log \big[ {\rm Tr} \rho_A^n \big] /(1-n)$, with $n=2$.
This is done through the ratio $ {\rm Tr} \rho_A^2= Z[A,2,T]/Z[A{=}0,2,T]$, where $Z[A{=}0,2,T] =  Z[T]^2$; the square of the unmodified partition function \cite{Melko:2010}.
One therefore needs to evaluate a difference in free energies or, alternatively, a ratio of partition functions,
between the replicated and unmodified simulation cells with QMC.
Several options are available, such as thermodynamic integration \cite{Melko:2010} and Wang-Landau sampling \cite{Inglis:2013}. Alternatively, methods such as the {\it ratio method} \cite{Gelman:1998,Humeniuk:2012,Luitz:2014}
allow one to compute the partition function ratio directly. As described in the next section, for this model, careful convergence to low temperature is required.  
Hence, in the Supplemental Material, we develop a highly-efficient variant of the ratio method, dubbed the {\it extended ensemble ratio method}, for the XY model.

{\em Results -- }
We perform large-scale calculations to obtain the second R\'enyi entropy of the XY model on
$L \times L$ square lattices with periodic boundary conditions.
The resulting torus is partitioned into two cylindrical regions of linear dimension $L \times L_x^A$ and $L \times (L-L_x^A)$.
For system sizes $L=8,12,16$, separate finite-temperature tests are performed to explore the convergence of $S_2$. In agreement with the expected scaling of
the tower-of-states gap, those results confirm that the convergence temperature scales approximately as $1/L^2$. 
We use this to estimate the convergence temperatures for $L=20,24,28,32$.

\begin{figure}[t]
\begin{center}
\scalebox{1}{\includegraphics[width=1.0\columnwidth]{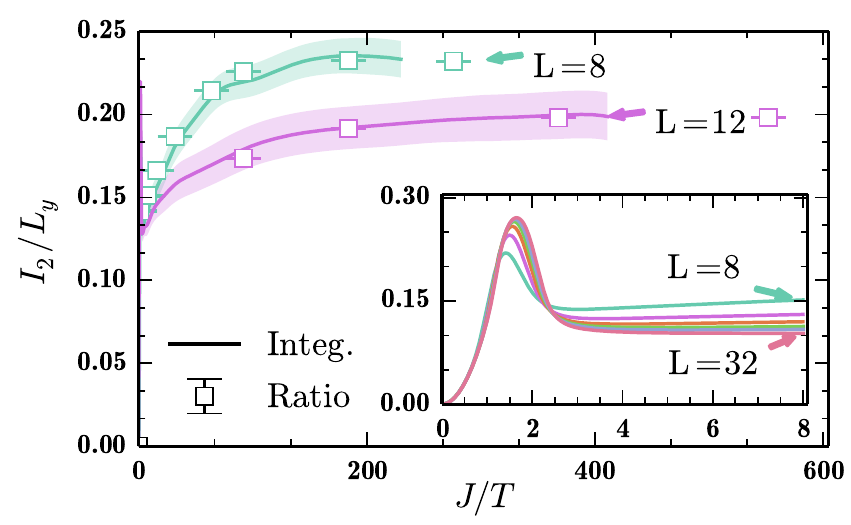}}
\end{center}
\caption{(color online).
The mutual information of the $S=1/2$ XY model as a function of $\beta \equiv J/T$.  Solid lines are obtained through thermodynamic integration from $\beta = 0$,
with statistical errors estimated by the shading.  Square points with error bars are data obtained at a fixed $\beta$ using the extended ensemble ratio method, 
described in the Supplemental Material.
}
\label{fig:conver}
\end{figure}

Since SSE QMC simulations necessarily run at finite temperature, 
very small thermal contributions to $S_2$ are expected, which we observe to
significantly affect our
finite-size scaling analysis below.
Fortunately, this thermal contribution can be
essentially eliminated by employing the mutual information, $I_2(A:B) = S_2(A) + S_2(B) - S_2(AB)$ \cite{Melko:2010}.
Assuming that bulk (volume-law) contributions from the two subsystems $A$ and $B$ approximately cancel the 
bulk term $S_2(AB)$ in $I_2$, the scaling form described by Metlitski and Grover becomes:
\begin{equation}
I_2 = a L + N_G \log (L \rho_s /c) + 2\gamma_{\rm ord}. \label{MI}
\end{equation}
Here, $a$ is a non-universal constant, $\rho_s$ and $c$ are the spin-wave stiffness and velocity, and $\gamma_{\rm ord}$ is 
the geometric constant that depends on the aspect ratios of the cylinders $A$ and $B$ \cite{Max_Tarun}.  Note, since non-universal 
(cutoff) dependences are all contained within $\rho_s$ and $c$, this geometric constant remains fully universal. 
For the spin-1/2 XY model on the square lattice, $\rho_s= 0.26974(5)J$ and $c= 1.1347(2)J$ were obtained from Ref.~\cite{Jiang:2011}.

Figure \ref{fig:conver} illustrates a representative convergence test for different system sizes.   The mutual information peaks at temperatures 
above the Kosterlitz-Thouless
transition of $(T/J)_{\rm KT} = 0.343$ (which can be detected by the crossing of the finite-size curves; see Ref.~\cite{Iaconis}).
For $T/J < (T/J)_{\rm KT}$, the mutual information reaches a minimum (at $J/T \equiv \beta \approx 4$ in Fig.~\ref{fig:conver}) before undergoing a slow rise. 
This rise continues until the approximate ground state is reached, for temperature below the finite-size scaling gap, which for system sizes larger than $L=8$ occurs for $ \beta > 100$.
Thus, although the method of thermodynamic integration is useful to produce the general shape of the $I_2$ curve for a wide range of temperatures, it is difficult to control 
the systematic error introduced by numerical integration at low temperatures for $L > 12$.  Therefore, data used in the below fits was converged at very low temperatures using the extended ensemble ratio 
method, described in the Supplemental Material.  

\begin{figure}
\begin{center}
\scalebox{1}{\includegraphics[width=1.0\columnwidth]{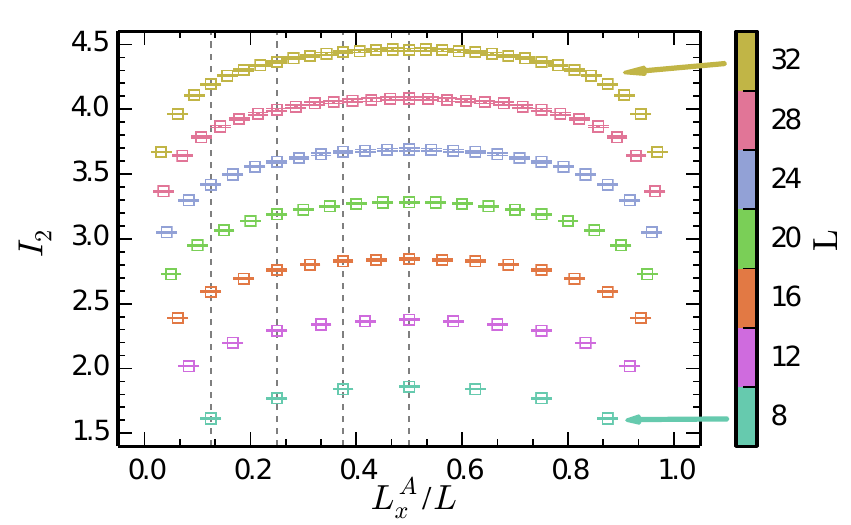}}
\end{center}
\caption{(color online).
The mutual information as a function of torus aspect ratio, for the lowest temperatures examined for each system size.
The corresponding $\beta$ are 184, 368, 736, 1150, 1650, 2300, 3200 ordered from the smallest to the largest system
size. Vertical dashed lines are the aspect ratio values employed in the fitting in Fig.~\ref{fig:fits2}
}
\label{fig:EEvsLv3}
\end{figure}

Figure \ref{fig:EEvsLv3} illustrates the resulting temperature-converged mutual information for a variety of system sizes, as a function of the
``width'' of the cylindrical region, $L_x^A$.  
Since, for a subsystem $A$ and its complement $B$, $S_A = S_B$ only at $T=0$,
the symmetry of the entanglement entropy about $L_x^A/L = 1/2$
provides a sensitive test of temperature convergence. 
Employment of the ``bare'' R\'enyi entropy results in a very slight asymmetry in the curve; use of $I_2$
restores this symmetry producing high-quality data that can be fitted using Eq.~(\ref{MI}).

\begin{figure}
\begin{center}
\scalebox{1}{\includegraphics[width=1.0\columnwidth]{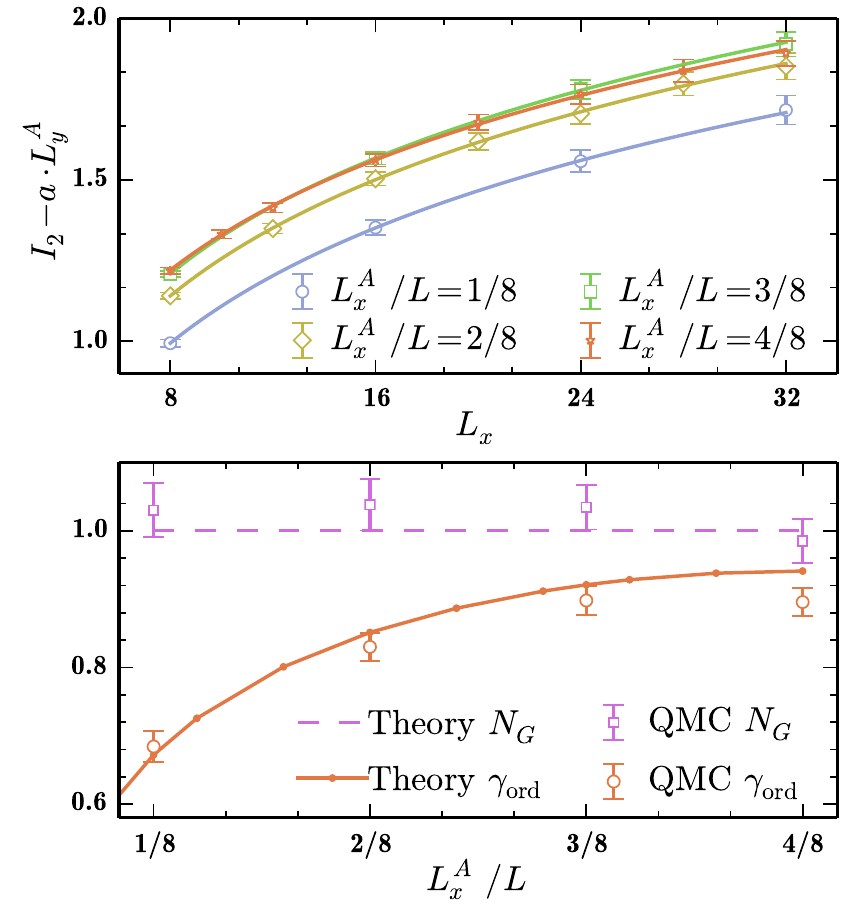}}
\end{center}
\caption{(color online). The top figure displays a three parameters fit to the 
functional form $a L+ b\log(L\rho_s/c)+d$ for different torus aspect ratios, where
$b$ gives a value for $N_{\rm G}$. 
Another three parameters fit to $aL +\log(L\rho_s/c)+2\gamma_{\rm ord}+d/L$ (not shown) is 
employed to extract the geometrical constant $\gamma_{\rm ord}$.  The resulting $N_{\rm G}$ and
$\gamma_{\rm ord}$ are shown alongside the theoretically-predicted values at bottom.} 
\label{fig:fits2}
\end{figure}

The results of this analysis are illustrated in Fig~\ref{fig:fits2}.  Here, $I_2$ is calculated at various aspect ratios
(the vertical cuts in Fig.~\ref{fig:EEvsLv3}) and fit to the functional form Eq.~(\ref{MI}).  Specifically, to extract the coefficient of the subleading
logarithm, the mutual information was fit to $I_2 = a L + b \log (L \rho_s /c) + d$, where $a,b$ and $d$ are adjustable fit parameters.   
As illustrated in  Fig.~\ref{fig:EEvsLv3} (a), there is definative evidence for the existence of a logarithm; furthermore, independent fits for 
the four aspect ratios studied each give $N_{\rm G} = 1$ to within error bars.  

Even more striking, we are able to extract the universal shape-dependence of the geometric constant $\gamma_{\rm ord}$.  To do so, fits were performed
to the functional form $I_2 = a L + \log (L \rho_s /c) + 2\gamma_{\rm ord} + d /L$, where $N_{\rm G}$ is fixed at unity in order to remove one parameter from the analysis.  
Thus calculated, $\gamma_{\rm ord}$ for $N=2$ in two dimensions can be compared via 
a zero-parameter fit to the subleading constant term $\gamma_{\rm free}$ calculated in a free scalar field theory \cite{Max_Tarun} through the relation
$\gamma_{\rm ord} = \gamma_{\rm free} + \frac{1}{2} \log(2\pi)$, valid for the second R\'enyi entropy.
The free field result $\gamma_{\rm free}$, which depends on the aspect ratio $L_x^A/L$, can be calculated numerically for free bosons on the lattice
using the correlation matrix technique (as in Ref.~\cite{Peschel:2009}).
As illustrated in Fig.~\ref{fig:EEvsLv3} (b), the resulting theoretical curve is in excellent agreement with our QMC results for $\gamma_{\rm ord}$.

{\em Discussion -- }
In this paper, we have employed large-scale quantum Monte Carlo (QMC) simulations on the spin-1/2 XY model to demonstrate the 
presence of a logarithmic correction to the R\'enyi entropy due to spontaneous breaking of continuous symmetry.
This term arises from the presence of two infra-red energy scales in the problem;
the spin-wave gap, and the ``tower of states''.
The coefficient of this logarithm is predicted in Ref.~\cite{Max_Tarun} to be $N_{\rm G}(d-1)/2$, where $N_{\rm G}$ is the number of Goldstone
modes and $d$ the spatial dimension.  We confirm this prediction in a striking manner, through finite-size scaling studies on the square lattice XY model,
recovering to high precision the expected $N_{\rm G} =1$.  In order to do so, a QMC algorithm was developed to sample the 
second R\'enyi entropy at a fixed temperature with high efficiency (described in the Supplemental Material).

Remarkably,  in addition to confirming  $N_{\rm G} =1$, our simulation technique is able to converge the value of an additional additive geometric
constant $\gamma_{\rm ord}$, which is fully universal since all short-distance physics is confined to the (known) spin wave stiffness and velocity, contained within the argument of the logarithm.  
The resulting $\gamma_{\rm ord}$ has a functional dependence on the geometric aspect ratio of the entangled bipartition. 
This function matches, to within error bars, that calculated using a free scalar field theory regularized on 
a toroidal square lattice, with no adjustable parameters.  

This is a rare and striking example of complete quantitative agreement between
a universal quantity calculated in continuum field theory and finite-size lattice simulations.
The ability of the R\'enyi entanglement entropy to mediate between continuum theory and lattice simulation 
illustrates the utility of such geometric quantities as an alternative paradigm to study correlations
in condensed-matter systems.
In this case, with the full understanding of the universal structure of the entanglement entropy in the 
presence of a spontaneously broken continuous symmetry, the door is now open to the direct search and detection
of Goldstone modes in a large variety of systems, through the widely accessible R\'enyi entropies.

\subsection*{Acknowledgments} 
We are thankful for enlightening discussions with A.~L\"auchli, D.~Poilblanc, K~Resch, and particularly, M.~Metlitski and T.~Grover, without whom this work would not have been possible.
Support was provided by NSERC, the Canada Research Chair program, and the Perimeter Institute (PI) for Theoretical Physics. Research at PI is supported by the Government of Canada through Industry Canada and by the Province of Ontario through the Ministry of Economic Development \& Innovation.  
S.I. acknowledges support from the FP7/ERC Starting Grant No. 306897.
The simulations were performed on the computing facilities of SHARCNET.

\putbib[Biblio]
\end{bibunit}

\begin{bibunit}[apsrev4-1]

\clearpage
\pagebreak
\begin{center}
\textbf{\large Supplementary material for ``Detecting Goldstone Modes with Entanglement Entropy''}
\end{center}

\setcounter{equation}{0}
\setcounter{figure}{0}
\setcounter{table}{0}
\setcounter{page}{1}
\makeatletter
\renewcommand{\theequation}{S\arabic{equation}}
\renewcommand{\thefigure}{S\arabic{figure}}
\renewcommand{\bibnumfmt}[1]{[S#1]}
\renewcommand{\citenumfont}[1]{S#1}

\section{Preliminaries}

This Supplemental Material describes a quantum Monte Carlo (QMC) algorithm
for estimating the R\'enyi entropy,
\begin{equation}
S_n = \frac{1}{1-n} \log \big[ {\rm Tr} \rho_A^n \big], \label{Renyi}
\end{equation}
using an extended-ensemble version of the so-called ``ratio method'' first applied in this 
context by Humeniuk and Roscilde \cite{Humeniuk:2012}.
In the present work, the algorithm will be specialized to the case of the second R\'enyi
entropy ($n=2$), although extensions to other $n$ are relatively straightforward.
Similarly, although we concentrate on applications to the spin-1/2 XY model,
generalizations to Hamiltonians with other symmetries are possible.

QMC estimators for the second R\'enyi entropy $S_2$ rely on the fact that the trace over
powers of the reduced density matrix can be related to a ratio of partition functions:
\begin{equation}
{\rm Tr} \rho_A^2= \frac{Z[A,2,T]}{Z[A{=}0,2,T]}, \label{Zratio}
\end{equation}
where the numerator is the partition function of a multi-sheeted Riemann surface \cite{Melko:2010}, 
(a ``replicated'' QMC simulation cell)
and the denominator is the square of the regular partition function, $Z[A{=}0,2,T] = Z[T]^2$. 
In the replicated case, the geometry of the entangled region $A$ dictates the ``boundaries''
of the $d+1$ dimensional QMC simulation cell; world-lines in region $A$ are periodic in 
imaginary time with period $2 \beta$ (or $n\beta$), while in region $B$, 2 (or $n$) independent replicas
exist with periodicity $\beta$.

Since the logarithm in Eq.~(\ref{Renyi}) reduces the calculation of $S_2$ to the difference in free 
energies between systems described by the two partition functions,
thermodynamic integration
or Wang-Landau techniques can be used to devise QMC estimators.  
In contrast, {\it ratio methods} (not to be confused with the related ``ratio trick'' coined in Ref.~\cite{Hastings:2010})
give an estimator for the ratio of partition functions, and act as valuable alternatives to
explicit calculation of free energies.
The most distinct advantage of ratio methods is their ability to calculate $S_n$ directly at a given 
fixed temperature.   They can however be inefficient for large entropies (large subregions $A$), and may
require several separate simulations of different subregion geometries (that can be combined with the ratio trick) to combat this.

\section{Extended-ensemble ratio method}

We begin by generalizing Eq.~(\ref{Zratio}) to the problem of calculation the ratio of replicated partition
functions defined with two arbitrary regions
$A$ and $A'$; namely $Z[A',2,T]/Z[A,2,T]$. 
As with all ratio methods, the QMC estimator in this case is based on a simple identity \cite{gelman1998}: 
\begin{align}
	\label{eq: simple_id}
	\frac{Z_{A'}}{Z_{A}} = \left \langle \frac{W_{A'}(c)}{W_A(c)} \right\rangle_{A},
\end{align}

\noindent 
where we have simplified our previous notation $Z_A = Z[A,2,T]$. The expectation 
value is taken in the $Z_A$ ensemble and $W_{A'} (c)$ and $W_{A}(c)$ are the weights 
of a configuration $c$ in the $Z_{A'}$ and $Z_A$ ensembles respectfully. 

The difficulty of applying this identity in a straightforward manner is that it 
is valid only when the configuration space of $Z_{A'}$, $\Omega_{A'}$, is 
contained within the configuration space of $Z_{A}$, $\Omega_{A}$ -- that is $W(c) \neq 0$ 
for any $c$.  Unfortunately, this condition is not automatically satisfied when the standard 
labelling variables are used in conventional QMC techniques. However, for the XY model, by introducing a new label,
it is possible to recast the configuration spaces of both ensembles such that 
$\Omega_{A}=\Omega_{A'}$. In other words, a common label $c$ can be found that 
enumerates all possible configurations within both ensembles.

\begin{figure}
\begin{center}
\scalebox{1}{\includegraphics[]{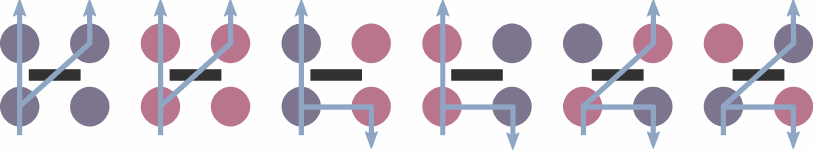}}
\end{center}
\caption{(color online). The six types of vertices present in a SSE simulation of 
the XY model \cite{Dir_loop}. The colored circles represent two possible states of a spin one-half. 
The horizontal solid bar depicts a two sites operator. The lines with arrow tips
show all possible non-bounce moves for a given vertex when the entrance leg is the 
left-most bottom one. There are two of those moves for each vertex.}
\label{fig: vertices}
\end{figure}

\begin{figure}
\begin{center}
\scalebox{1}{\includegraphics[]{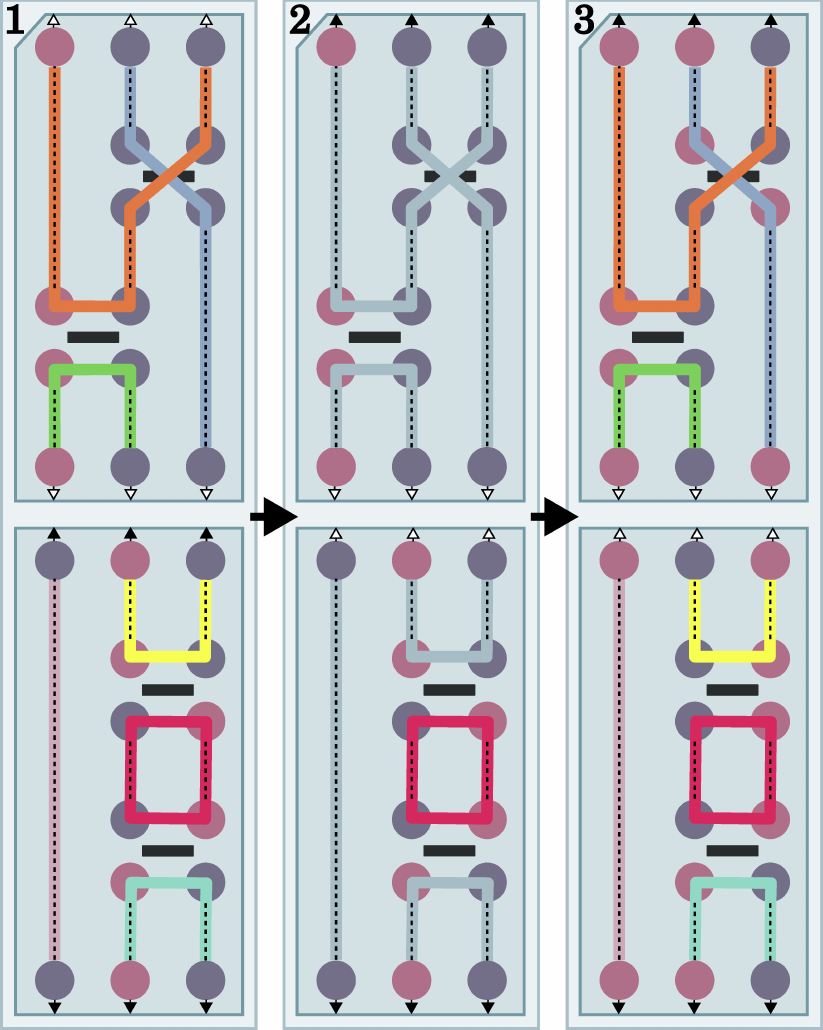}}
\end{center}
\caption{(color online). A three step conversion process of a vertex configuration 
$v^A(l)$ into $v^{A'}(l)$ that preserves its segment partition. The details of 
this process are illustrated in the text. At each step, a simulation cell composed 
of two replicas (top and bottom) is shown. The dashed vertical lines between adjacent
legs that are located on different vertices form a linked list. 
The label $l$ identifies different sets of those lines. The two types of small 
arrows placed next to the replicas' boundary slices mark the boundary conditions 
along the dimension of the expansion: within a column, spins  decorated with the 
same kind of arrows are connected.  In this way, the first simulation cell's 
region $A$ is empty while the other two simulation cells' region $A'$ contains all 
three spins.  Open colored solid lines trace out a segment partition of the first 
and the third simulation cells (7 segments total). Note that within the same cells, 
there is also a single closed segment, an inner loop, composed of four legs. In 
the second simulation cell, the open segments are merged by boundary connections 
to form a single cross-replica loop identified by the same color. Mismatching
boundary spins along this loop are flipped according to the algorithm presented
in the text, resulting into a vertex configuration compatible with the 
new boundary conditions as displayed in the third simulation cell.
}
\label{fig: cells}
\end{figure}

Here we will consider a stochastic series expansion (SSE) representation of the 
partition function~\cite{SSE1,SSE2,Sandvik:1999}. A term in the SSE expansion of $Z$ can 
be represented by a spin state and an operator list, where the operators 
are terms in the Hamiltonian. Alternatively, we can represent the same configuration 
by a linked list of vertices \cite{Dir_loop}; an example of a set of vertices is displayed
in Fig.~\ref{fig: vertices}. We can therefore label the configuration by the list of links, $l$, and the set of vertex types $v^A(l)$ 
for a given boundary condition defined by $A$ as shown in Fig.~\ref{fig: cells}. 
Here we have defined $l$ to include 
all ``internal'' links between vertices and we link the ``exterior'' vertices to the boundaries 
of the operator string; we specifically do not close the links at the boundaries, as one
would do in a regular $d+1$ dimensional QMC simulation cell \cite{Dir_loop}

These two labels are sufficient to identify the configuration space of many models. 
Thus, a partition function can be expressed as the following double sum:
\begin{align}
\label{eq: gen_part_func}
	Z_A = \sum_{l} \sum_{v^A(l)} W(v^A(l)),
\end{align}
where $W(v^A(l))$ is the weight of a configuration labelled by $l$ and $v^A(l)$.

This general expression can be simplified for the XY model. The model's SSE vertices 
are displayed in Fig.~\ref{fig: vertices}. With an appropriate choice of adjustable 
SSE constants, the weight of each vertex becomes equal to $1/2 J$ \cite{Dir_loop}. Hence, the weight 
of a vertex configuration does not depend on a particular combination of 
vertices in the list and, therefore, is completely defined by the length of the 
corresponding operator list alone. We get that $W_{XY}(v^A(l))=W_{XY}(l)$ and 
Eq.~(\ref{eq: gen_part_func}) is simplified to: 
\begin{align}
\label{eq: XY_part_func}
	Z_A^{XY}=\sum_{l} W_{XY}(l) \sum_{v^A(l)} 1.
\end{align}

The second sum counts the degeneracy of vertex configurations compatible with 
the boundary conditions between replicas. Motivated by the search for a variable 
independent of those boundary conditions, we introduce a new label, $s(v^A(l))$, 
that enumerates all possible partitions of a vertex configuration labelled by 
$(l,v^A(l))$ into a set of non-overlapping segments. The following algorithm is 
used to construct a single instance of those segments:
\begin{enumerate}
	\item Pick an unmarked leg located on a boundary slice. Mark it as visited.
	\item By following the linked list, switch to a leg connected to it.
	\item The new leg belongs to a vertex. 
		  \begin{itemize}
			  	\item If this vertex is unmarked, pick with an equal probability 
					  one of two possible non-bounce moves for this vertex and 
					  switch to the corresponding leg.  Mark this vertex as 
					  visited and store the move type.
				\item If the vertex is marked, switch to the next leg by performing 
				      a move of the same type that was done before.
		  \end{itemize}
	\item Repeat steps 2-4 until a leg on a boundary slice is reached.
\end{enumerate}
By repeating this algorithm for all legs located on the boundary slices, all open 
segments are traced out. However, it is possible that some of the legs located
on the inner slices have been left unmarked after this procedure. In order to 
partition those remaining legs too, the  closed segments (loops) need to be traced.  
This is achieved by adjusting two steps of the algorithm. Now in the first step, 
the choice of legs to be picked is extended to all interior legs.  Once the initial 
leg is picked, the algorithm proceeds in the same way until it reaches the same 
leg again. Hence, the condition to terminate the execution of the forth step has 
to be modified appropriately. By construction, any two segments built in such a 
manner can never pass through the same leg and, therefore, are non-intersecting.

The loops tracing continues until all legs are marked. By the end of this procedure 
every leg belongs to one single segment (closed or open). This constitutes a single
instance of the partitioning of a vertex configuration into a set of non-overlapping 
segments. An example of such partition is shown in the first cell of Fig.~\ref{fig: cells}.
Note that at each vertex, there are two choices how to proceed with the construction 
of a segment.  Each of them leads to a different partition.\footnote{Two partitions are 
considered equal when all their segments are the same. In its turn, for two 
segments to be considered the same the order of legs in the construction of one 
segment must match exactly with the order of legs in the construction of another 
segment} Therefore, a simulation cell that contains $N_v$ vertices can be partitioned 
in $2^{N_v}$ distinct ways. This is the range of the newly introduced label $s$. 
Since $N_v$ just counts the number of vertices without discerning their types, 
the number of partitions for a particular $v^A(l)$ is determined by the $l$ label only.
This fact allows to rewrite Eq.~(\ref{eq: XY_part_func}) in a new form:
\begin{align}
\label{eq: XY_part_func_2}
	Z_A^{XY}=\sum_{l}  \frac{W_{XY}(l)}{2^{N_v(l)}} \sum_{v^A(l)}   \sum_{s(v^A(l))} 1 .
\end{align}
This expression is obtained from Eq.~(\ref{eq: XY_part_func}) by introducing 
a third sum via the substitution: $1=\frac{1}{2^{N_v(l)}} \sum_{s(v^A(l))} 1$ where
the new sum is performed over all different partitions of $v^A(l)$.

It can be also shown that for any vertex configuration in $A$, $v^A(l)$, partitioned 
as $s(v^A(l))$, there exists a vertex configuration in $A'$, ${v'}^{A'}(l)$, 
with exactly the same partitioning, that is  $s({v'}^{A'}(l)) = s(v^A(l))$. The proof is 
by construction.  If $v^A(l)$ and ${v'}^{A'}(l)$ were the same, the task is trivial. 
Otherwise, $v^A(l)$ has to be modified in order to satisfy the boundary conditions 
of $A'$. An example of such process is displayed in Fig.~\ref{fig: cells}.  Here, 
the first and thirds simulation cells represent $v^A(l)$ and ${v'}^{A'}(l)$
correspondingly. The second cell depicts an intermediate step of the correctional
procedure. Here, the open segments are connected into a loop along which the boundary
spins mismatches are fixed one-by-one. Further details of the algorithm are given 
below.

Proceeding column by column, consider each pair of boundary legs, $(s_1^0,s_2^0)$, 
to be matched  with respect to the new boundary conditions $A'$. If the legs align,
$s_1^0=s_2^0$, proceed to the next pair. Otherwise, randomly choose one of the two legs 
in the pair. Say it is $s_1^0$. Since this leg is located on a boundary slice, it belongs to an 
open segment. Flip all legs belonging to this segment. Now, the original pair 
of legs is properly aligned, however there might be another mismatch at the other 
end of the segment. Call the new pair $(s_1^1,s_2^1)$ where $s_2^1$ is the leg that 
has just been flipped as part of the open segment. By the same logic as before,
$s_1^1$ must belong to an open segment whose other end is identified as another
boundary leg $s_2^2$.  If $s_1^1 \neq s_2^1$,  flip this segment in order to align
the $(s^1_1,s^1_2)$ pair and move on to the next pair $(s^2_2,s^2_1)$. Otherwise, 
proceed to the same pair without flipping the segment. In this way, one-by-one 
pairs of boundary legs are aligned with respect to $A'$ boundary condition along 
a loop of open segments. An important subtlety occurs at the last step of this 
algorithm when the last pair $(s^n_2, s^1_2)$ is considered. Unlike previously, 
$s^1_2$ cannot be flipped if those legs do not align. An attempt to do so would 
entail another iteration of corrections with the same result, thus, initiating 
the algorithm in an infinite loop.

However, this does not occur in the XY-model due to the special properties of its 
vertices. Notice that the only vertex move that connects two anti-aligned legs
is the ``switch-and-reverse'' move (Fig.\ref{fig: vertices}); this is the only 
move that reverses the vertical directionality of propagation of the segment's 
head. Consequently, once a segment tracing is initiated with the choice of a leg 
and its state, the spin state of the leg at the segment's head is determined 
by the vertical direction that the segment passes through the leg.  On the last 
boundary connection, the direction of motion along the segment must be the same 
as the initial direction, and therefore  the initial spin state at the head of 
the segment under construction is always the same its final state.  We see then 
that for any segment partition of $v^A(l)$, it is always possible to construct 
a ${v'}^{A'}(l)$ with the same segment partition.

Now that we have shown the equivalence between any two configuration spaces constraint
by boundary conditions $A$ and $A'$ in terms of the segment partitions, we have
achieved our initial goal to find a label $s$ that can be used to apply the Eq.~\ref{eq: simple_id}. 
However, we still need to compute the weights of $W^{A}(s)$ 
and $W^{A'}(s)$. Upon a close inspection of the inner double sum in Eq.~\ref{eq: XY_part_func_2}, 
we realize that the same segment partitions can be generated from different 
vertex configurations. This degeneracy can be exploited in order to replace
the double sum with a single sum:
\begin{align}
\label{eq: XY_part_func_3}
	Z_A^{XY}=\sum_{l}  \frac{W_{XY}(l)}{2^{N_v(l)}} \sum_{s(l)} \mathrm{deg}^A(s(l)) ,
\end{align}
where the inner sum iterates over all unique segment partitions for a given linked
list and $\mathrm{deg}^A(s(l))$ is the degeneracy of the segment partition labelled as $s(l)$.
As will be shown, the degeneracy depends on the boundary conditions between replicas
and, thus, the superscript must be included. 

In order to calculate the degeneracy of a segment partition $s(l)$, connect 
open segments using boundary conditions $A$ to form $N_{b}^A(s(l))$ loops that 
cross boundary slices. In addition to those loops, there are also $N_{i}(s(l))$ 
inner loops (closed segments). Since all those loops do not intersect with each 
other by construction, the spins within them can be flipped independently. Each 
combination of the loops' flips leads to another vertex configuration. 
Equivalently, all of those vertex configurations generate the same segment 
partition $s(l))$. There are in total $N_{b}^A(s(l))+N_{i}(s(l))$ loops with each 
one being in one of two states: flipped or not-flipped. Each combination of those
states corresponds to a different vertex configuration. In total, there are 
$2^{N_{b}^A(s(l))+N_{i}(s(l))}$ of such combinations which constitutes the degeneracy 
$\mathrm{deg}^A(s(l))$.

With the last piece of the puzzle in our hands, we infer the segment partition 
weight from Eq.~(\ref{eq: XY_part_func_3}):
\begin{align}
\label{eq: weight}
	W(s(l))= \frac{W_{XY}(l)}{2^{N_v(l)}} 2^{N_{b}^A(s(l))+N_{i}(s(l))}.
\end{align}
Since the inner-loops are unaffected by the boundary conditions, upon the substitution
of this weight in the Eq.~(\ref{eq: simple_id}) their number drops out and an elegant 
expression is obtained:
\begin{align}
\label{eq: estimator}
	\frac{Z_{A'}}{Z_{A}} = \langle 2^{N_b^{A'}(s(l))- N_b^A(s(l))} \rangle_{A}.
\end{align}

In practice, this estimator can be implemented in fewer steps that were required
to prove its validity. It requires two routines. One routine traces out a random 
single segment partition $s(l)$ for a vertex configuration $v^A(l)$ as was 
outlined before.  In order to speed up the execution, it is not necessary to 
identify the closed segments. The end product of this routine is to associate the 
pairs of boundary spins that are connected via open segments. Once this step is 
done, the second routine takes the set of those pairs together with replicas' 
boundary conditions as its inputs. Its task is to count $N_b^{A}$. This routine 
is executed for both $A$ and $A'$ with the same open segments. In the end, 
$N_b^{A'}(s(l))$ and $N_b^{A}(s(l))$ are known and the estimator can be evaluated
according to Eq.~(\ref{eq: estimator}).

\section{Benchmarks}
\begin{figure}[!ht]
\begin{center}
\scalebox{1}{\includegraphics[width=\columnwidth]{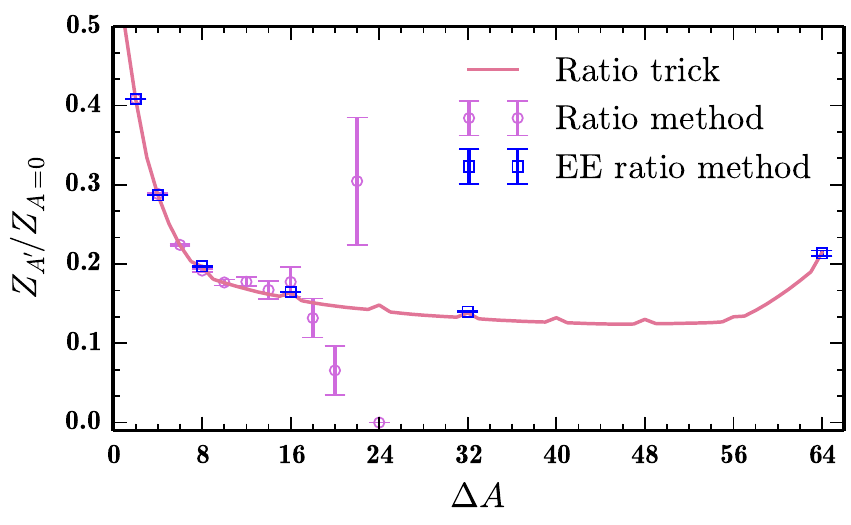}}
\end{center}
\caption{The comparison of ratio methods efficiencies in a $8\times 8$ system with
periodic boundary conditions at $\beta=8$. The ratio of partition functions measurement 
is plotted against the size difference between their corresponding regions A, 
$\Delta A$. The ratio method completely fails for $\Delta A > 22$, and thus the
data is not shown on the plot. The values obtained from the ratio trick serve as a reference.  The 
statistical error in those values is contained within the width of the curve.} 
\label{fig:methcomp}
 \end{figure}

To illustrate the efficiency of the new estimator, its raw measurements are compared 
to the original ratio method \cite{Humeniuk:2012} in Fig.~\ref{fig:methcomp}, on the 2D
XY model of interest in the main text.
Here, the deterioration of both 
estimators' statistics is seen as the difference between the partition functions'
regions A, $\Delta A {=} A'-A$, grows large. 
As the reference values, we employ the results 
obtained from the ratio trick \cite{Hastings:2010}, which constitute a compilation of the extended ensemble 
(EE) ratio method results from many different Monte Carlo simulations, each executed 
with $\Delta A{=}1$. Note that the original ratio method results are based on five 
times more Monte Carlo sweeps that were involved to produce the EE ratio method 
results. Even with such advantage, the ratio method statistics becomes increasingly
poor towards $\Delta A {=} 22$. After this threshold, the estimator seems no longer 
capable to capture any meaningful statistics within the running time of its simulations.
The performance of the EE is strikingly better. Even when $\Delta A {=} 64$, the largest
possible increment in the system, it produces an accurate result with a precision
comparable to the ratio method precision at $\Delta A {=} 10$.

\putbib[Biblio]
\end{bibunit}
\nopagebreak

\end{document}